\documentclass{article}

\usepackage{arxiv}

\usepackage[utf8]{inputenc} 
\usepackage[T1]{fontenc}    
\usepackage{hyperref}       
\usepackage{url}            
\usepackage{booktabs}       
\usepackage{amsfonts}       
\usepackage{nicefrac}       
\usepackage{microtype}      
\usepackage{lipsum}
\usepackage{amsmath}
\usepackage{float}
\usepackage{subfigure}
\usepackage{graphicx}
\usepackage{tikz}
\usepackage{bm}
\usetikzlibrary{arrows}
\usepackage{adjustbox}
\graphicspath{ {./images/} }
\DeclareUnicodeCharacter{FB01}{fi}

\title{Segmentation of carotid vessel wall using U-Net and segmentation average network}

\author{
 Mingjie Jiang \\
  Department of Electrical Engineering\\
  City University of Hong Kong\\
  Hong Kong \\
  \texttt{minjiang5-c@my.cityu.edu.hk} \\
   \And
 J. David Spence \\
  Stroke Prevention \& Atherosclerosis Research Centre\\
  Robarts Research Institute\\
  London, Ontario, Canada\\
  \texttt{dspence@robarts.ca} \\
  \And
 Bernard Chiu \\
  Department of Electrical Engineering\\
  City University of Hong Kong\\
  Hong Kong \\
  \texttt{bcychiu@cityu.edu.hk} \\
}

\begin{document}
\maketitle
\begin{abstract}
Segmentation of carotid vessel wall is required in vessel wall volume (VWV) and local vessel-wall-plus-plaque thickness (VWT) quantiﬁcation of the carotid artery. Manual segmentation of the vessel wall is time-consuming and prone to interobserver variability. In this paper, we proposed a convolutional neural network to segment the common carotid artery (CCA) from 3D carotid ultrasound images. The proposed CNN involves three U-Nets that segmented the 3D ultrasound (3DUS) images in the axial, lateral and frontal orientations. The segmentation maps generated by three U-Nets were consolidated by a novel segmentation average network (SAN) we proposed in this paper. The experimental results show that the proposed CNN improved the segmentation accuracies. Compared to only using U-Net alone, the proposed CNN improved the Dice similarity coefficient (DSC) for vessel wall segmentation from $64.8\%$ to $67.5\%$, the sensitivity from $63.8\%$ to $70.5\%$, and the area under receiver operator characteristic curve (AUC) from 0.89 to 0.94. 
\end{abstract}


\section{Introduction}
Stroke is the second global leading cause of death~\cite{world2019global}. Atherosclerosis is a major cause of ischemic stroke. Fortunately, for patients with high stroke risk, lifestyle, dietary and medical therapies reduce the occurrence of stroke by $75-80\%$~\cite{cheng2017sensitive}. With the advances in the pathogenesis of atherosclerosis, new therapeutic targets and corresponding treatments are expected to be developed. In parallel to the development of new treatment strategies, there is a critical need for sensitive biomarkers that are able to detect treatment effects in clinical trials.

Ainsworth et al.~\cite{ainsworth20053d} reported that 3D Carotid Ultrasound (3DUS) Total Plaque volume (TPV) is a sensitive way to measure effects of atorvastatin on atherosclerosis. Krasinski et at.~\cite{krasinski2009three} reported that carotid 3DUS vessel wall volume (VWV) is a biomarker sensitive to the effect of atorvastatin therapies. Cheng et al.~\cite{cheng2017sensitive} showed that carotid vessel-wall-plus-plaque thickness (VWT) measured from 3DUS is sensitive to B-Vitamin treatments, which are expected to confer a smaller beneficial effect than medical therapies, such as statins. Segmentation of vessel wall is required in VWV and VWT quantiﬁcation of the carotid artery. However, manual segmentation of vessel wall is time-consuming and prone to interobserver variability. Therefore, there is a need for efficient and automatic methods for segmenting the carotid vessel wall, bounded by the media-adventitia (MAB) and lumen-intima (LIB).

Deep-learning segmentation models, such as U-Net~\cite{ronneberger2015u,dong2017automatic,oktay2018attention,zhou2018unet++,kohl2018probabilistic}, have achieved high performance in biomedical image segmentation. In particular, Zhou et al.~\cite{zhou2019deep} have used U-Net to segment axial slices of 3DUS images. The method required a human observer to identify a number of points lying on the MAB, from which an initial MAB was generated. At each point on the initial MAB, image patches were generated by sliding a window along the normal of the initial MAB. These image patches were then fed to a dynamic convolutional neural network (CNN) to obtain the final MAB contour. The common carotid artery (CCA) images were then cropped based on the segmented MAB contour, from which the LIB contours were segmented using the U-Net. This method requires 13.2s for a human observer to initialize the MAB. Considering 25 slices are typically needed for a clinical study~\cite{egger2007validation}, the time required from a human observer is high in a clinical trial involving hundreds of patients, such as the risk stratification study in~\cite{wannarong2013progression,van2014three}. The goal of this study was to develop a fully automatic vessel wall segmentation method. Since 3DUS images were reconstructed from contiguous 2D axial images acquired by the ultrasound machine~\cite{fenster19963} and the in-plane resolution is higher than the elevational resolution, carotid segmentation is typically performed on the axial images either manually~\cite{egger2007validation} or semi-automatically~\cite{zhou2019deep,ukwatta2010sci}. In addition to segmenting the axial plane, we propose in this paper to segment the carotid arteries on two perpendicular longitudinal views, which we call lateral and frontal views. The segmentation results obtained on these three perpendicular planes were consolidated by a CNN proposed in this paper, which we call segmentation average network (SAN). In this work, we aim at validating the hypothesis that this approach provides more accurate segmentation results than U-Net.

\section{Method}
\label{sc:me}
\subsection{Image preprocessing}
The proposed algorithm was evaluated on 3DUS images of the CCA acquired for 22 subjects with diabetic nephropathy. The subjects were recruited from the Nephrology Clinics and Diabetes Clinics at the London Health Science Centre (London, Canada), and they provided written informed consent to the study protocol approved by a local research ethics board~\cite{cheng2017sensitive}. High-resolution 3DUS images were obtained by translating an ultrasound tranducer (L12-5, Philips, Bothel, WA, USA) mounted on a mechanical assembly along the neck of the subjects for approximately 4 cm. The 2D ultrasound frames from the ultrasound machine (ATL HDI 5000, Philips, Bothel, WA, USA) were captured by a frame grabber and reconstructed into a 3D image.   

For the implementation of segmentation networks, we resize all the axial, lateral and frontal images into $320\times 256$, $320\times 128$ and $256\times 128$, respectively, then we prepared three separate training sets $\mathcal{D}_x$, $\mathcal{D}_y$ and $\mathcal{D}_z$, consisting of the axial, lateral and frontal slices, respectively. The number of slices used for training and testing for $\mathcal{D}_x$, $\mathcal{D}_y$ and $\mathcal{D}_z$ are shown in Table~\ref{table:num_train_test}.
\begin{table}[b]
\centering
\caption{Number of slices for training and testing}
\label{table:num_train_test}
\begin{adjustbox}{max width=\textwidth}
\begin{tabular}{cccc}

& $\mathcal{D}_x$ & $\mathcal{D}_y$ & $\mathcal{D}_z$\\
\hline
Training & 5428 & 5917 & 7305\\
Testing & 1387 & 1602 & 2169 \\
\end{tabular}
\end{adjustbox}
\end{table}

Fig~\ref{fig:slices} shows example axial, lateral and frontal slices for the common carotid artery of a subject. The resliced images were rescaled to $[-1, 1]$ as follows:
\begin{equation*}
\begin{aligned}
I(i,j) &\leftarrow  2\bigg(\frac{I(i,j)-\min{(I)}}{\max{(I)}-\min{(I)}}-0.5\bigg),
\end{aligned}
\end{equation*}
where $\min(\cdot)$ and $\max(\cdot)$ denotes the minimal and maximal pixel values of an image $I$, respectively.
\newcommand{\iical}{0.19\textwidth}
\begin{figure*}[thbp]
\subfigure[axial]{
\includegraphics[width=\iical]{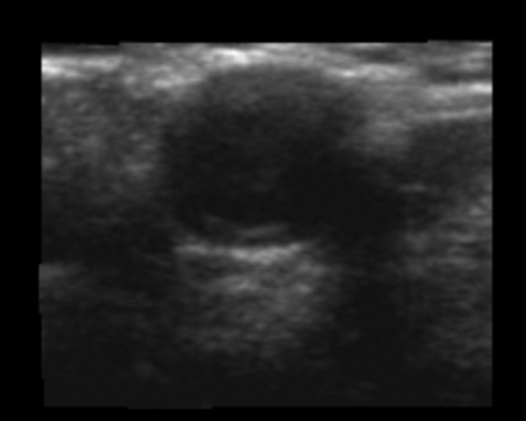}}
\hfill
\subfigure[lateral]{
\hfill
\includegraphics[width=\iical]{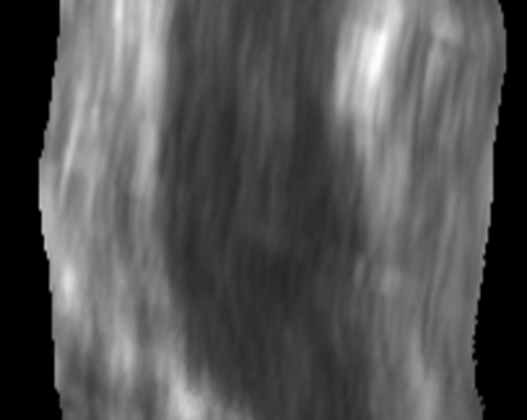}}
\hfill
\subfigure[frontal]{
\includegraphics[width=\iical]{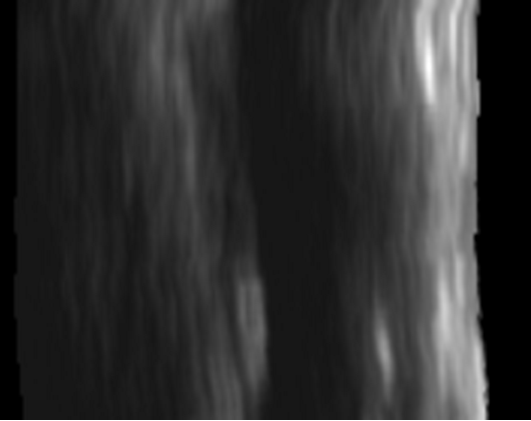}}
\caption{\label{fig:slices}Axial, lateral and frontal slices of CCA 3DUS}
\end{figure*}

\subsection{Network architecture and training}
\begin{figure*}[h]
\centering
\includegraphics[width=0.69\textwidth]{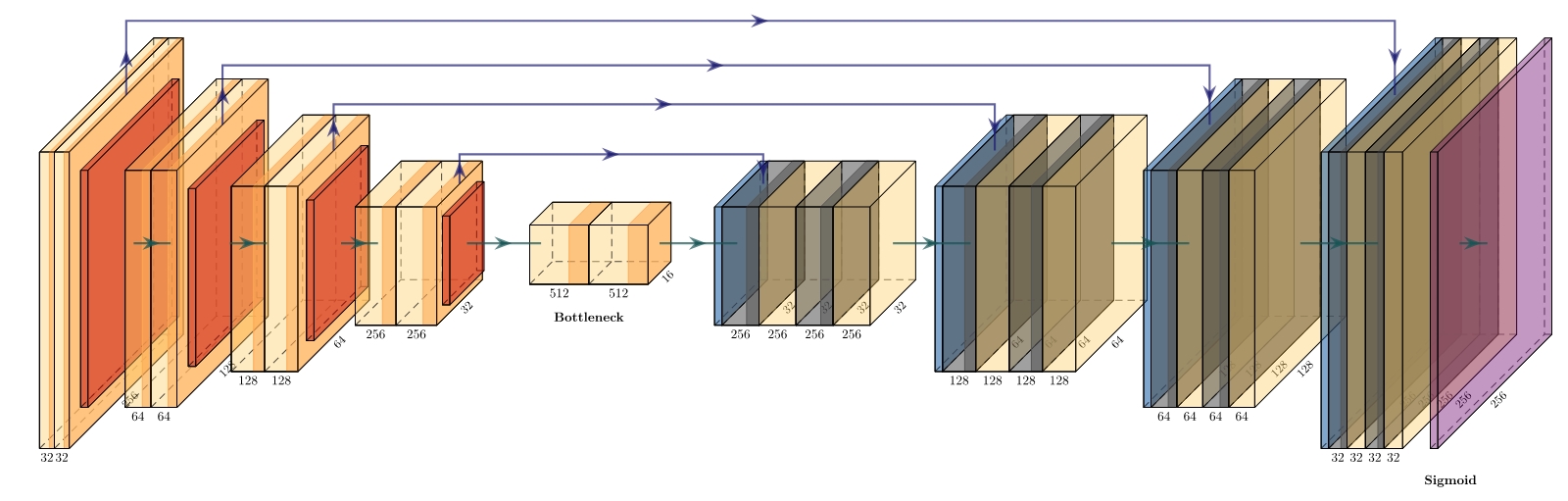}
\caption{U-Net architecture\label{fig:unet}}
\end{figure*}

Our segmentation model is based on U-Net~\cite{ronneberger2015u} and batch normalization~\cite{ioffe2015batch}, which is added between convolution and activation function. Fig~\ref{fig:unet} shows the U-Net architecture used in this study.

The stochastic optimization tool Adam~\cite{kingma2014adam} was used to train our networks and the related hyperparameters setting are given as follows: $\alpha=10^{-3}$, $\beta_1=0.9$, $\beta_2=0.999$, $\epsilon=10^{-7}$, batch size $m=16$, number of epochs $n_\text{epoch}=50$.

The loss function in training is based on DSC and mean-squared difference between the algorithm and manually segmented results:
$$\mathcal{L}(\widehat{s}, s)=-\frac{2\sum\widehat{s}_{ij}s_{ij}}{\sum(\widehat{s}_{ij}+s_{ij})}+\frac{1}{N}\sum(\widehat{s}_{ij}-s_{ij})^2,$$
where $N$ is the number of pixels; $\widehat{s}_{ij}$, $s_{ij}$ refer to the segmentation label generated by the algorithm and manual segmentation, respectively.

\begin{figure}[thbp]
\centering
\includegraphics[width=0.39\textwidth]{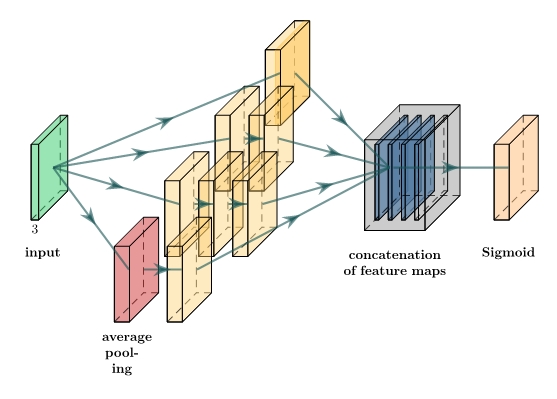}
\caption{SAN architecture\label{fig:inception}}
\end{figure}

We built three U-Net models trained with $\mathcal{D}_x$, $\mathcal{D}_y$ and $\mathcal{D}_z$ separately, and we call these models $\mathcal{M}_x$, $\mathcal{M}_y$ and $\mathcal{M}_z$. After we trained these models, we built a small CNN, segmentation average network (SAN), to average the segmentation results of $\mathcal{M}_x$, $\mathcal{M}_y$ and $\mathcal{M}_z$. The architecture of SAN is based on an inception block~\cite{szegedy2015going} shown in Fig~\ref{fig:inception}, where yellow block denotes convolution layer and red block represents average pooling. 

\subsection{Inference}
\begin{figure}[thbp]
\begin{center}
\resizebox{0.34\textwidth}{!}{%
\begin{tikzpicture}[ 
font=\sffamily, 
every matrix/.style={ampersand replacement=\&,column sep=0.5cm,row sep=0.5cm}, source/.style={draw,thick,rounded corners,fill=yellow!20,inner sep=.3cm}, process/.style={draw,thick,circle,fill=blue!20}, sink/.style={source,fill=green!20}, rectangle/.style={draw,very thick,shape=rectangle,inner sep=.3cm}, dots/.style={gray,scale=2}, to/.style={->,>=stealth',shorten >=1pt,semithick, font=\sffamily\footnotesize}, every node/.style={align=center}]  

\matrix{ \node[process] (reslice1) {reslice}; \& 
\node[rectangle] (buffer1) {$\mathcal{M}_y$}; \&
\node[process] (reslice2) {reslice}; \\
\node[source] (hisparcbox) {3D image}; \& 
\node[rectangle] (buffer2) {$\mathcal{M}_x$}; \& 
\node[rectangle] (buffer5) {SAN}; \&
\node[sink] (datastore) {output}; \\ 
\node[process] (reslice3) {reslice}; \& 
\node[rectangle] (buffer3) {$\mathcal{M}_z$}; \&
\node[process] (reslice4) {reslice}; \\};

\draw[to] (hisparcbox) --  (buffer2); 
\draw[to] (hisparcbox) --  (reslice1); 
\draw[to] (reslice1) --  (buffer1); 
\draw[to] (buffer1) --  (reslice2); 
\draw[to] (reslice2) --  (buffer5); 
\draw[to] (hisparcbox) --  (reslice3);
\draw[to] (reslice3) --  (buffer3); 
\draw[to] (buffer3) --  (reslice4); 
\draw[to] (reslice4) --  (buffer5);
\draw[to] (buffer2) --  (buffer5);
\draw[to] (buffer5) --  (datastore);

\end{tikzpicture}}%
\end{center}
\caption{\label{fig:predict_SAN}Inference with U-Net$+$SAN}
\end{figure}
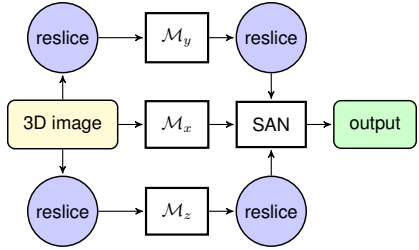

Fig~\ref{fig:predict_SAN} shows how we combined the segmentation results generated by $\mathcal{M}_x$, $\mathcal{M}_y$ and $\mathcal{M}_z$. First, we resliced a 3DUS image into axial, lateral and frontal slices, respectively. Then we input axial slices into $\mathcal{M}_x$ and input lateral and frontal slices into $\mathcal{M}_y$ and $\mathcal{M}_z$. Finally, we resliced the binary segmentation volume generated by $\mathcal{M}_y$ and $\mathcal{M}_z$ so that the resliced binary maps have the same orientation as that generated by $\mathcal{M}_x$. Three different binary segmentation maps in the axial orientation were available after the reslicing operations. These three maps were fed to SAN to obtain the final segmentation map for each axial image.

\section{Experimental results}
\label{sc:ex}
We used Python 3.7 as programming language, and employed the deep learning framework Keras~\cite{chollet2015keras} with TensorFlow~\cite{tensorflow2015-whitepaper} as a backend. We trained our CNN models with an {\sc NVIDIA RTX2080 Ti GPU}. The training and segmentation time is shown in Table~\ref{table:time}.
\begin{table}[b]
\centering
\caption{Average training and segmentation time}
\label{table:time}
\begin{adjustbox}{max width=\textwidth}
\begin{tabular}{ccc}
model& Total training time & segmentation time (per slice) \\
\hline
U-Net (axial) & {$5750s$} & $0.084s$ \\
U-Net (lateral) & {$3350s$} & {$0.016s$} \\
U-Net (frontal) & {$3300s$}   & {$0.013s$} \\
SAN & $1750s$ & $0.041s$ \\
\end{tabular}
\end{adjustbox}
\end{table}

We quantified segmentation performance of our method by DSC, sensitivity and Intersection-Over-Union (IoU) defined below: $\text{DSC}(\widehat{s}, s) = \frac{2\sum\widehat{s}_{ij}s_{ij}}{\sum(\widehat{s}_{ij}+s_{ij})}$, $\text{Sensitivity}(\widehat{s}, s) =\frac{\sum\widehat{s}_{ij}s_{ij}}{\sum s_{ij}}$, $\text{IoU}(\widehat{s}, s) = \frac{\sum\widehat{s}_{ij}s_{ij}}{\sum(\widehat{s}_{ij}+s_{ij})-\sum\widehat{s}_{ij}s_{ij}}$.

The experimental accuracies are given in Table~\ref{table:conclusion}. Some segmentation results are given in Fig~\ref{fig:unet_bound}.
\begin{table}[b]
\centering
\caption{Accuracies}
\label{table:conclusion}
\begin{adjustbox}{max width=\textwidth}
\begin{tabular}{cccc}

model& DSC & Sensitivity & IoU\\
\hline
U-Net & {$64.8\%$}     &$63.8\%$ & $51.1\%$\\
U-Net$+$SAN & \bm{$67.5\%$}     &\bm{$70.5\%$} &\bm{$53.0\%$} \\
\end{tabular}
\end{adjustbox}
\end{table}
\newcommand{\iiical}{0.19\textwidth}
\begin{figure*}[thbp]
\subfigure[original]{
\includegraphics[width=\iiical]{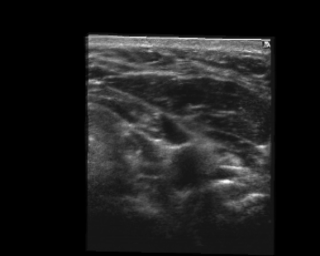}}
\hfill
\subfigure[manual segmentation]{
\includegraphics[width=\iiical]{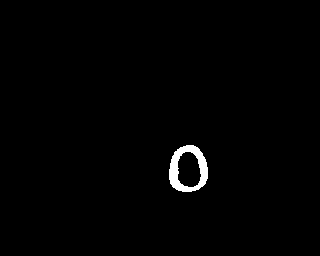}}
\hfill
\subfigure[U-Net]{
\includegraphics[width=\iiical]{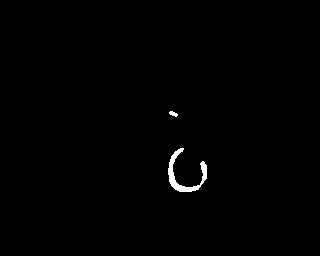}}
\hfill
\subfigure[U-Net$+$SAN]{
\includegraphics[width=\iiical]{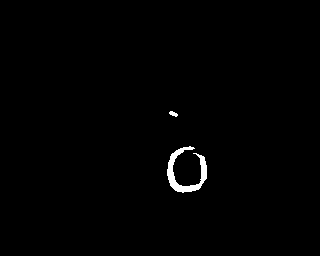}}
\caption{\label{fig:unet_bound}Segmented US images by U-Net and U-Net$+$SAN}
\end{figure*}
Since the sigmoid function was used as the activation function in the output layer, each pixel value of the output is between 0 and 1. We varied the threshold between 0 and 1 to compute the true positive rate and the false positive rate, and generated the receiver operating characteristic (ROC) curves of U-Net and U-Net+SAN, which are shown in Fig~\ref{fig:roc_auc}. The area-under-curve (AUC) of U-Net$+$SAN (0.94) is higher than the AUC of U-Net (0.89).

\begin{figure}[thbp]
\centering
\includegraphics[width=0.39\textwidth]{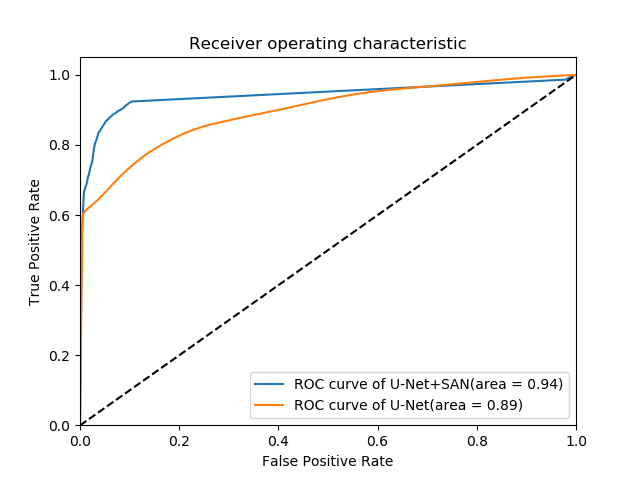}
\caption{ROC curves and their AUC\label{fig:roc_auc}}
\end{figure}

\section{Discussion and Conclusion}
\label{sc:co}
    In this paper, we used three U-Nets to segment the axial, lateral and frontal slices of 3DUS CCA images and proposed a very small CNN, which we call SAN, to consolidate the segmentation results. This method can incorporate the information of the slices of three directions of 3D images amd provide higher segmentation accuracy than solely using U-Net in the axial images. 
    
  The most substantial advantage of our fully automatic method as compared against semi-automatic segmentation method is that no human interaction time is needed. Compared to Zhou et al.~\cite{zhou2019deep} and Ukwatta et al.~\cite{ukwatta2013three}, which requires ~10s for a user to initialize contours on an axial slice, the proposed fully automated method is more suitable for clinical trials that involves evaluation of hundreds of patients~\cite{wannarong2013progression,van2014three}. Although the DSC we reported in this study is around $70\%$ compared to over $90\%$ in Zhou et al.~\cite{zhou2019deep} and Ukwatta et al.~\cite{ukwatta2013three}, we would emphasize that the DSC is quantified for the vessel wall here (i.e., the region between the LIB and MAB contours), whereas the DSC is quantified separately for the LIB and MAB contours in Zhou et al and Ukwatta et al.. The two results are not comparable as DSC penalizes contours with smaller area more heavily. Given the same distance error between a larger and a smaller object (such as the distance error synthetically generated by rigid translating the gold standard boundary for a fixed distance), DSC is higher for the larger object as the area overlap (expressed as the percentage of the entire region covered by either the algorithm or manual segmentation) is higher for the larger object. When evaluated separately for MAB and LIB, the DSC attained by our algorithm is also over $90\%$. Another advantage of the proposed method is that we segmented the vessel wall directly, which guarantees that the lumen lies completely inside the MAB. There is no such guarantee for methods segmenting MAB and LIB separately, such as Zhou et al. and Ukwatta et al. However, a limitation of the proposed algorithm is that it does not guarantee a closed vessel wall (e.g., Fig. 6). The next version of our algorithm will guarantee closed vessel walls. 
\section*{Acknowledgement}
Dr. Chiu is grateful for funding support from the Research Grant Council of the HKSAR, China (Project nos. CityU 11205917, CityU 11203218) and the City University of Hong Kong Strategic Research Grants (nos. 7004617, 7005226).

\bibliographystyle{unsrt}  
\bibliography{main}  


\end{document}